\documentclass[onecolumn,journal]{IEEEtran}

\usepackage{graphicx}
\usepackage{float}
\usepackage{amsmath}
\usepackage{subfigure} 
\usepackage{mathtools}
\usepackage{amssymb} 
\usepackage{color}
\usepackage{soul}
\usepackage{booktabs}

\newcommand{\Droe}{{\bf D}_{\sigma_e}}
\newcommand{\Drom}{{\bf D}_{\sigma_m}}
\newcommand{\De}{{\bf D}_{\epsilon}}
\newcommand{\Dm}{{\bf D}_{\mu}}
\newcommand{\K}{\bf K}

\newcommand{\scroe}{{\sigma_e}}
\newcommand{\scrom}{{\sigma_m}}
\newcommand{\sce}{{\epsilon}}
\newcommand{\scm}{{\mu}}

\newcommand{\wDroe}{\widetilde{\bf D}_{\sigma_e}}
\newcommand{\wDrom}{\widetilde{\bf D}_{\sigma_m}}
\newcommand{\wDe}{\widetilde{\bf D}_{\epsilon}}
\newcommand{\wDm}{\widetilde{\bf D}_{\mu}}
\newcommand{\wK}{\widetilde{\bf K}}

\newcommand{\vect}[1]{\mathbf{#1}}
\newcommand{\matr}[1]{\mathbf{#1}}

\newcommand{\pref}[1]{(\ref{#1})}

\begin{document}
\title{Structure-Preserving Reduction of Finite-Difference Time-Domain Equations with Controllable Stability Beyond the CFL Limit}

\author{Xihao~Li,~\IEEEmembership{Student~Member,~IEEE,}
        Costas~D.~Sarris,~\IEEEmembership{Senior~Member,~IEEE,}
        and~Piero~Triverio,~\IEEEmembership{Member,~IEEE}\\
        Submitted to IEEE Transactions on Microwave Theory and Techniques on June 25, 2014
\thanks{This work was supported in part by the Natural Sciences and Engineering Research Council of
Canada (Discovery grant program) and in part by the Canada Research Chairs program.}
\thanks{X. Li, C. D. Sarris and P.~Triverio are with the Edward S. Rogers Sr. Department of Electrical and Computer Engineering, University of Toronto, Toronto, M5S 3G4 Canada (email: xihao.li@mail.utoronto.ca, costas.sarris@utoronto.ca, piero.triverio@utoronto.ca).}
}

\markboth{}
{}

\maketitle

\begin{abstract}
The timestep of the Finite-Difference Time-Domain method (FDTD) is constrained by the stability limit known as the Courant-Friedrichs-Lewy (CFL) condition. This limit can make FDTD  simulations quite time consuming for structures containing small geometrical details. Several methods have been proposed in the literature to extend the CFL limit, including implicit FDTD methods and filtering techniques. In this paper, we propose a novel approach which combines model order reduction and a perturbation algorithm to accelerate FDTD simulations beyond the CFL barrier. We compare the proposed algorithm against existing implicit and explicit CFL extension techniques, demonstrating increased accuracy and performance on a large number of test cases, including resonant cavities, a waveguide structure, a focusing metascreen and a microstrip filter.
\end{abstract}


\IEEEpeerreviewmaketitle

\section{Introduction}

The Finite-Difference Time-Domain method (Yee's FDTD) is one of the most popular algorithms for solving Maxwell's equations~\cite{FDTD}. Standard FDTD updates electric and magnetic field values with a leap-frog scheme which ensures second-order accurate approximations of time and spatial derivatives. A remarkable feature of FDTD is that it avoids expensive matrix inversions due to an explicit discretization of time derivatives.  As a consequence, FDTD timestep is constrained by the Courant-Friedrichs-Lewy (CFL) stability condition~\cite{FDTD}
\begin{equation}
	{\Delta t} \le \frac{1}{c \sqrt{\frac{1}{\Delta x^2} + \frac{1}{\Delta y^2} + \frac{1}{\Delta z^2}}}\,,
	\label{eq:CFL}
\end{equation}
where $c$ is the wave velocity in the medium and $\Delta x$, $\Delta y$, $\Delta z$ denote the cell size in the three dimensions. Limit~\pref{eq:CFL} dictates a very small timestep in problems containing small geometric features relative to the wavelength. In such cases, both the large number of unknowns and the small timestep can make FDTD simulations very time consuming.

The CFL limit can be overcome in several ways. Implicit methods are unconditionally stable for any timestep, but require matrix inversions~\cite{FDTD_NO_TIMESTEPPING,FDTD_NO_TIMESTEPPING_2,Cle99}. The direct use of implicit techniques is prohibitive even for medium-size problems. Model order reduction (MOR) has been used to reduce the computational complexity of an implicit approach in combination with subgridding~\cite{Subcell,SecondOrder}. The Alternating-Direction-Implicit FDTD (ADI-FDTD)~\cite{ADI-FDTD,ADI-FDTD-2} has been proposed in an attempt to maintain some of the efficiency of explicit FDTD while ensuring unconditional stability.  ADI-FDTD splits the time stepping process into implicit and explicit half steps, thus guaranteeing unconditional stability~\cite{ADI-FDTD}.  This step incurs additional computation costs, although partially mitigated through the use of a larger timestep.

Recently, alternative methods to overcome the CFL limit have been proposed. In spatial filtering~\cite{SpatialFiltering}, unstable harmonics that arise above the CFL limit are removed at runtime using a Fast Fourier Transform and low-pass filtering. This approach is simple to implement, but requires special care at material boundaries to avoid aliasing in the Fourier transform process. In~\cite{DJiao,DJiao2014}, stable FDTD simulations beyond the CFL limit have been obtained by first running a short FDTD simulation at a stable timestep, in order to identify the dominant and stable eigenmodes of the structure. This information is then used to remove the unstable modes from the FDTD equations and run above the CFL limit.

In this paper, we propose a new way to accelerate FDTD simulations beyond the CFL limit, combining model order reduction and eigenvalue perturbation. Model order reduction is applied to FDTD equations to reduce their order. For a timestep \emph{below} the CFL limit, we prove that the reduced model is stable by construction. For timesteps \emph{above} the CFL limit, we propose a perturbation algorithm to enforce late time stability. The reduction process and the extension of the stability limit make the proposed technique faster than standard FDTD, with negligible loss of accuracy. Moreover, the proposed reduction process preserves the structure of the original FDTD equations, making the reduced model easy to integrate in an existing FDTD code. Some preliminary results from the proposed approach were presented, without proofs, in~\cite{APS}.

The paper makes the following two contributions:
\begin{enumerate}
\item we show how the CFL limit can be extended without having to perform time-consuming filtering operations at runtime, as in spatial filtering~\cite{SpatialFiltering}, or having to identify the dominant eigenmodes of the structure with a pre-processing step that requires careful monitoring of convergence~\cite{DJiao,DJiao2014};
\item we propose a model order reduction method for FDTD equations that works directly in the discrete time domain, as opposed to previous methods that worked in the continuous time domain~\cite{Subcell,SecondOrder,FDTD_NO_TIMESTEPPING}. This approach makes the enforcement of late time stability straightforward, and preserves the structure of FDTD equations, which was not maintained in our previous method~\cite{IMS}.
\end{enumerate}

The paper is organized as follows. In Sec.~\ref{sec:system}, we cast FDTD equations in matrix form and develop the proposed reduction algorithm. We also prove that the reduced model is stable by construction for timesteps below the CFL limit. Sec.~\ref{sec:stabilityenforcement} shows how stability can be enforced for timesteps \emph{above} the CFL limit. In Sec.~\ref{sec:implementation}, we discuss implementation details. Finally, in Sec.~\ref{sec:results} we apply the proposed method to five test structures, demonstrating its excellent accuracy, computational efficiency, and scalability to 3-D problems of practical interest.

\section{Stability-Preserving Model Order Reduction of FDTD Equations}
\label{sec:system}

\subsection{Matrix Formulation of FDTD Equations}

We start from Yee's FDTD equations~\cite{FDTD} that, in the one-dimensional case\footnote{For the sake of readability, we state FDTD equations in one dimension. All results presented in the paper are valid in the general 3-D case, as well as in 1-D and 2-D.}, read
\begin{subequations}
\begin{flalign}
&\left(\frac{\sce}{\Delta t} + \frac{\scroe}{2} \right) {E}|^{n+1}_{k} = \label{eq:FDTD1} \\
&\;\;\;\;\;\;\;\;\;\;\;\left(\frac{\sce}{\Delta t} - \frac{\scroe}{2} \right)  {E}|^{n}_{k} - 
{\frac{1}{\Delta z}} \left({ H}|^{n+\frac{1}{2}}_{k+\frac{1}{2}} - {H}|^{n+\frac{1}{2}}_{k-\frac{1}{2}}\right) -J|^{n+\frac{1}{2}}_k 
\nonumber \\
&\left(\frac{\scm}{\Delta t} + \frac{\scrom}{2} \right){H}|^{n+\frac{3}{2}}_{k+\frac{1}{2}} -
{\frac{1}{\Delta z}} \left({E}|^{n+1}_{k+1} - {E}|^{n+1}_{k}\right) = \label{eq:FDTD2} \\
&\;\;\;\;\;\;\;\;\;\;\;\left(\frac{\scm}{\Delta t} - \frac{\scrom}{2} \right) {H}|^{n+\frac{1}{2}}_{k+\frac{1}{2}} - M|^{n+1}_{k+\frac{1}{2}} \nonumber \,.
\end{flalign}
\end{subequations}
In the equations above, ${E}|^{n}_{k}$ denotes the electric field at time $n$ and position $k$, while ${H}|^{n+\frac{1}{2}}_{k+\frac{1}{2}}$ denotes the magnetic field. With $\varepsilon$, $\mu$, $\sigma_e$ and $\sigma_m$, we denote permittivity, permeability, electric conductivity, and magnetic conductivity, respectively. Terms $J|^{n+\frac{1}{2}}_k$ and $M|^{n+1}_{k+\frac{1}{2}}$ denote electric and magnetic sources. 
FDTD equations can be arranged into matrix form~\cite{Denecker}
\begin{align}
&\begin{bmatrix}
\frac{{\De}}{\Delta t} + \frac{{\Droe}}{2} & 0 \\
-{\K}^{T} & \frac{{\Dm}}{\Delta t} + \frac{{\Drom}}{2}
\end{bmatrix}
\begin{bmatrix}
{\bf E}|^{n+1} \\
{\bf H}|^{n+\frac{3}{2}}
\end{bmatrix}
= \nonumber \\
&\;\;\;\;\;\;
\begin{bmatrix}
\frac{{\De}}{\Delta t} - \frac{{\Droe}}{2} & -{\K} \\
0 & \frac{{\Dm}}{\Delta t} - \frac{{\Drom}}{2}
\end{bmatrix}
\begin{bmatrix}
{\bf E}|^{n} \\
{\bf H}|^{n+\frac{1}{2}}
\end{bmatrix}
 + {\bf B}{\bf u}|^{n+1}
 \label{eq:matrix}
\end{align}
where:
\begin{itemize}
\item ${\De}$ and ${\Dm}$ are diagonal matrices containing the electric permittivity and magnetic permeability value for each cell;
\item ${\Droe}$ and ${\Drom}$ are diagonal matrices containing the electric and magnetic conductivity values for each cell;
\item matrix ${\K}$ arises from the discretization of the curl operators in Maxwell's equations, and contains terms in the form ($\pm 1/\Delta x$, $\pm 1/\Delta y$, $\pm 1/\Delta z$);
\item  vector ${\bf u}|^{n+1}$ includes all sources;
\item matrix ${\bf B}$ contains 1's corresponding to electric and magnetic source locations.
\end{itemize}
Representation~\pref{eq:matrix} holds for general 3D problems, with non-uniform material properties and a non-uniform Cartesian grid~\cite{Denecker}.
To compact the notation, we rewrite~\pref{eq:matrix} as~\cite{Denecker}
\begin{equation}
({\bf R} + {\bf F}){\bf x}|^{n+1} = ({\bf R} - {\bf F}){\bf x}|^{n} + {\bf B}{\bf u}|^{n+1}\,, \\
\label{MNAform}
\end{equation}
with
\begin{equation}
{\bf R}  = 
\begin{bmatrix}
       \frac{1}{\Delta t}{\De} & -\frac{1}{2} {\K} \\[2pt]
       -\frac{1}{2} {\K}^T  & \frac{1}{\Delta t}{\Dm}  \\
\end{bmatrix}\;\;\;\;\;
{\bf F}  = 
\begin{bmatrix}
       \frac{1}{2}{\Droe} & \frac{1}{2} {\K} \\
       -\frac{1}{2} {\K}^T  & \frac{1}{2}{\Drom}  \\
\end{bmatrix}\,,
\label{R_and_F}
\end{equation} 
and where
\begin{equation}
{\bf x}|^{n} = 
\begin{bmatrix}
       {\bf E}|^{n} \\ {\bf H}|^{n+\frac{1}{2}}
\end{bmatrix}
\label{eq:x}
\end{equation}
is a vector of size $N = N_e + N_h$, where $N_e$ is the number of electric field unknowns, and $N_h$ is the number of magnetic field unknowns.

\subsection{Stability Conditions}

Written in form~\pref{MNAform}, a system of FDTD equations can be interpreted as a discrete time system with input ${\bf u}|^{n+1}$ and state ${\bf x}|^{n}$. Its solution will be stable if and only if all poles of~\pref{MNAform} are inside the unit circle in the complex plane~\cite{Opp83} or, equivalently,  if the following two conditions hold 
\begin{eqnarray}
{\bf F}^T + {\bf F} &=& 
\begin{bmatrix}
       {\Droe} & 0 \\
       0  & {\Drom}  \\
\end{bmatrix}
\ge 0 \label{Dstability1} \\
{\bf R} &=&
\begin{bmatrix}
       \frac{1}{\Delta t}{\De} & -\frac{1}{2} {\K} \\[2pt]
       -\tfrac{1}{2} {\K}^T  & \frac{1}{\Delta t}{\Dm}  \\
\end{bmatrix} 
> 0 \label{Dstability2}
\end{eqnarray}
where $\ge 0$ denotes a positive semidefinite matrix, and $>0$ denotes a positive definite matrix\footnote{A symmetric matrix $\matr{A}$ is positive definite if for any vector $\vect{x} \neq 0$ we have $\vect{x}^T \matr{A} \vect{x} > 0$. It is positive semidefinite if  $\vect{x}^T \matr{A} \vect{x} \ge 0$.}.
Conditions~\pref{Dstability1} and~\pref{Dstability2} were proposed in~\cite{Denecker} and have an intuitive physical explanation. Inequality~\pref{Dstability1} simply requires all conductivities to be positive.  Inequality~\pref{Dstability2} can be shown~\cite{Denecker} to be equivalent to the CFL limit~\pref{eq:CFL} and limits the maximum timestep $\Delta t$ that can be used in a stable FDTD simulation.

\subsection{Model Order Reduction}

We now reduce FDTD equations~\pref{MNAform} using the SPRIM model order reduction technique~\cite{SPRIM}. Firstly, from the matrices in~\pref{MNAform}, we generate a projection matrix 
\begin{equation}
{\bf V} = 
\begin{bmatrix}
       {\bf V}_1 & 0 \\
       0 & {\bf V}_2 
\end{bmatrix}
\label{eq:V}
\end{equation}
using the robust Arnoldi process~\cite{SPRIM, PRIMA}. Matrices ${\bf V}_1$ and ${\bf V}_2$ are orthonormal and of size $N_e \times \widetilde{N}$ and $N_h \times \widetilde{N}$, respectively, with $\widetilde{N}$ much smaller than $N_e$ and $N_h$. Then, we approximate the full vector of unknowns $\vect{x}|^n$ with a \emph{reduced} vector $\widetilde{\vect{x}}|^n$ as
\begin{equation}
\vect{x}|^n
\simeq
{\bf V}\;\widetilde{\bf x}|^{n+1}
\label{eq:projection}
\end{equation}
Substituting~\pref{eq:projection} into~\pref{MNAform}, and multiplying on the left by $\matr{V}^T$, we obtain
\begin{equation}
{\bf V}^T({\bf R} + {\bf F}) {\bf V} \;\; \widetilde{\bf x}|^{n+1} = {\bf V}^T({\bf R} - {\bf F}) {\bf V} \;\; \widetilde{\bf x}|^{n} + {\bf V}^T{\bf B}{\bf u}|^{n}\\
\label{reducedMNAform1}
\end{equation}
and, after carrying out matrix multiplications,
\begin{equation}
(\widetilde{\bf R} + \widetilde{\bf F}) \widetilde{\bf x}|^{n+1} = (\widetilde{\bf R} - \widetilde{\bf F}) \widetilde{\bf x}|^{n} + \widetilde{\bf B}{\bf u}|^{n}\,,\\
\label{reducedMNAform}
\end{equation}
where $\widetilde{\bf R} = {\bf V}^T{\bf R} {\bf V}$, $\widetilde{\bf F} = {\bf V}^T{\bf F} {\bf V}$ and $\widetilde{\bf B} = {\bf V}^T{\bf B}$ are ``compressed'' versions of $\matr{R}$, $\matr{F}$ and $\matr{B}$, respectively. The order of~\pref{reducedMNAform} is $2\widetilde{N}$, which is much lower than then order $N$ of the original system~\pref{MNAform}. The reduced model order and, consequently, its accuracy, can be controlled by choosing the number of columns $\tilde{N}$ of the projection matrices $\matr{V}_1$ and $\matr{V}_2$ generated by the Arnoldi algorithm. Due to the small size, \pref{reducedMNAform} can be solved very quickly to find the reduced unknowns $\widetilde{\bf x}|^{n}$.
Once $\widetilde{\bf x}|^{n}$ is available, the fields at any point in the system are computed through~\pref{eq:projection}.

Using~\pref{R_and_F} and~\pref{eq:projection}, the matrices in~\pref{reducedMNAform} can be written as
\begin{equation}
\widetilde{\bf R}  = 
\begin{bmatrix}
       \frac{1}{\Delta t}{\wDe} & -\frac{1}{2} {\wK} \\[2pt]
       -\frac{1}{2} {\wK}^T  & \frac{1}{\Delta t}{\wDm}  \\
\end{bmatrix}\;\;\;\;\;
\widetilde{\bf F}  = 
\begin{bmatrix}
       \frac{1}{2}{\wDroe} & \frac{1}{2} {\wK} \\[2pt]
       -\frac{1}{2} {\wK}^T  & \frac{1}{2}{\wDrom}  \\
\end{bmatrix}\,,
\label{R_and_Freduced}
\end{equation} 
where
\begin{align}
{\wDe} &= {{\bf V}_1}^T {\De}{\bf V}_1 &
{\wDm}    &= {{\bf V}_2}^T {\Dm}{\bf V}_2 \\ 
{\wDroe} &= {{\bf V}_1}^T {\Droe}{\bf V}_1 &
{\wDrom} &= {{\bf V}_2}^T {\Drom}{\bf V}_2 \\
{\wK}            &= {{\bf V}_1}^T {\K}{\bf V}_2
\end{align}
Owing to the block-diagonal nature of the projection matrix~\pref{eq:V} used in  SPRIM~\cite{SPRIM}, the reduction process preserved the structure of the original FDTD equations~\pref{MNAform}, which is a novel result. Being in the same form as FDTD equations, the solution of~\pref{reducedMNAform} can be computed in a leap-frog manner, for increased efficiency. 

\subsection{Stability Preservation Below the CFL Limit}

We now discuss the stability of the reduced model~\pref{reducedMNAform}. First, we consider the case when $\Delta t$ is below the CFL limit of the original FDTD equations, showing that the obtained reduced model is stable \emph{by construction}. Since we have preserved the structure of the FDTD equations, stability conditions~\pref{Dstability1} and~\pref{Dstability2} can be also applied to the reduced model
\begin{align}
{{\bf \widetilde{F}}}^T + {{\bf \widetilde{F}}} \ge 0 \label{rDstab1}\\
  {\bf \widetilde{R}} > 0 \label{rDstab2}
\end{align}
The first condition can be rewritten as
\begin{equation}
{{\bf \widetilde{F}}}^T + {{\bf \widetilde{F}}} = 
\begin{bmatrix}
       {\bf V}_1 & 0 \\
       0 & {\bf V}_2 
\end{bmatrix}^T 
\left({\bf F}^T + {\bf F}\right)
\begin{bmatrix}
       {\bf V}_1 & 0 \\
       0 & {\bf V}_2 
\end{bmatrix} \ge0\,.
\label{rDstability1}
\end{equation}
Since the original model~\pref{MNAform} satisfies~\pref{Dstability1}, the last expression in~\pref{rDstability1} is positive semidefinite by construction, as it is the congruence of a positive semidefinite matrix~\cite{Hor12}. Similarly, since $\matr{R} > 0$ because of~\pref{Dstability2}, and $\matr{V}$ is full rank, we have 
\begin{equation}
{\bf \widetilde{R}} = 
\begin{bmatrix}
       {\bf V}_1 & 0 \\
       0 & {\bf V}_2 
\end{bmatrix}^T 
{\bf R}
\begin{bmatrix}
       {\bf V}_1 & 0 \\
       0 & {\bf V}_2 
\end{bmatrix}>0
\end{equation}
Therefore, the proposed approach preserves stability by construction, avoiding the need for an additional post-processing step to enforce its stability as in~\cite{IMS}.  

\section{Stability Enforcement Above the CFL Limit}
\label{sec:stabilityenforcement}

\subsection{Theoretical Derivation}
\label{sec:enforce}

If the chosen $\Delta t$ is beyond the CFL limit of the original FDTD equations, conditions~\pref{Dstability1} and~\pref{Dstability2} will be violated and reduced model~\pref{reducedMNAform} may contain unstable eigenvalues. However, due to its small size, its stability can be easily enforced, effectively breaking the CFL barrier.

From stability criteria~\pref{Dstability1} and~\pref{Dstability2}, we see that changing $\Delta t$ will only affect the second condition, since the first one does not depend on $\Delta t$. In order to make the reduced model stable, we need to enforce
\begin{equation}
{\bf \widetilde{R}} = 
\begin{bmatrix}
       \frac{1}{\Delta t}{\wDe} & -\frac{1}{2} {\wK} \\
       -\frac{1}{2} {{\wK}}^T  & \frac{1}{\Delta t}{\wDm}  \\
\end{bmatrix} 
> 0\,, \label{rDstability2}
\end{equation}
which can be achieved by perturbing $\wK$ as follows. Using the Schur complement~\cite{Boy94}, we can state two conditions equivalent to~\pref{rDstability2}
\begin{align}
\frac{1}{\Delta t}{\wDm} & > 0 \label{stability2} \\
\frac{1}{\Delta t}{\wDe} & -  \frac{\Delta t}{4} \wK \wDm^{-1} \wK^T > 0 \label{stability3}
\end{align}
It can be seen that~\pref{stability2} always holds, while~\pref{stability3} is the only source of potential instability at refined CFL numbers. Rearranging terms in~\pref{stability3}, we arrive at the following inequality
\begin{equation}
\begin{aligned}
({{\wDe}}^{-\frac{1}{2}} {\wK} {{\wDm}}^{-\frac{1}{2}}) ({{\wDe}}^{-\frac{1}{2}} {{\wK}} {{\wDm}}^{-\frac{1}{2}})^T < \frac{4}{{\Delta t}^2}{\bf I}\,.
\label{stability4}
\end{aligned}
\end{equation}
If we denote  the singular values~\cite{Hor12} of ${{\wDe}}^{-\frac{1}{2}} {\wK} {{\bf \widetilde{D}}_{\mu}}^{-\frac{1}{2}}$ as $\sigma_i$ for $i = 1,\dots,\tilde{N}$, we have that~\pref{stability4} holds if and only if~\cite{Denecker}
\begin{equation}
\sigma_{i} < \frac{2}{\Delta t} \quad \text{for } i = 1,\dots,\tilde{N}\,.
\label{singular_limit}
\end{equation}
Above the CFL limit, some singular values $\sigma_i$ may violate~\pref{singular_limit}, and make the reduced model unstable. In order to enforce its stability, we propose the following procedure:
\begin{enumerate}
\item Compute the singular value decomposition~\cite{Hor12}
\begin{align*}
{{\wDe}}^{-\frac{1}{2}} {\wK} {{\bf \widetilde{D}}_{\mu}}^{-\frac{1}{2}} = {\bf U} {\bf S} {\bf W}^T\,,
\end{align*}
where $\matr{S}$ is a diagonal matrix containing the singular values $\sigma_i$. This operation is cheap since it is performed on a small matrix of size $\tilde{N} \times \tilde{N}$.
\item Perturb the singular values $\sigma_i$ which exceed~\pref{singular_limit}
	\begin{equation}
		\sigma'_i = \begin{cases}
					\sigma_i &\text{if } \sigma_i < \gamma \frac{2}{\Delta t}  \\
					\gamma \frac{2}{\Delta t} &\text{otherwise}
					\end{cases}
	\end{equation}
	where $\gamma$ is slightly less than one. In the examples of Sec.~\ref{sec:results}, we used $\gamma = 0.9999$.
	Form a new diagonal matrix $\matr{S}'$ with the perturbed singular values $\sigma'_i$.
\item Obtain the perturbed ${\wK}^\prime$ matrix  
\begin{align*}
&{\wK}^\prime = {{\wDe}}^{\frac{1}{2}}{\bf U}{\bf S}^\prime{\bf W}^T{{\bf \widetilde{D}}_{\mu}}^{\frac{1}{2}}
\end{align*}
and replace $\wK$ with $\wK'$ in~\pref{R_and_Freduced}. 
\end{enumerate}
This procedure leads to a reduced model which satisfies~\pref{rDstab1} and~\pref{rDstab2} by construction, and is thus stable for a timestep above the CFL limit. With the proposed technique, the CFL limit can be extended without having to switch to an implicit formulation or perform filtering operations at runtime~\cite{SpatialFiltering}, which reduce computational efficiency.

\subsection{Demonstration of Stability Enforcement}
\label{sec:demo}

\begin{figure}[t]
\centering
{\includegraphics[scale=1]{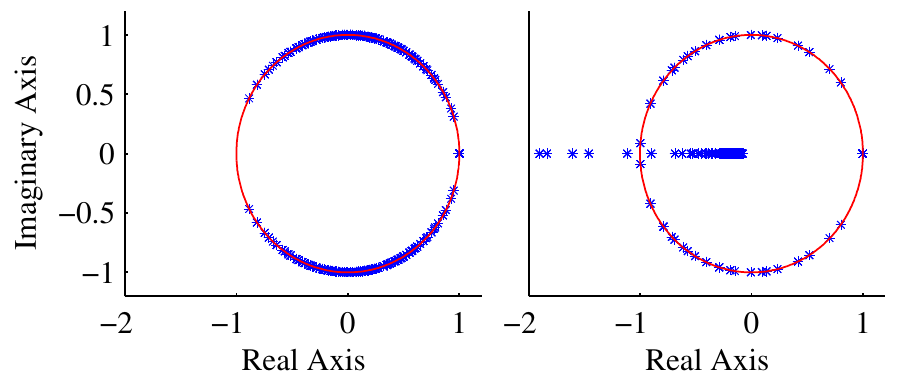}}
\caption{Example of Sec.~\ref{sec:demo}: eigenvalues of FDTD equations~\pref{MNAform} below the CFL limit (left panel, $s=0.99$) and above the CFL limit (right panel, $s=1.98$). The stability region is given by the red circle.}
\label{demo_fig1}
\end{figure}

\begin{figure}[t]
\centering
{\includegraphics[scale=1,trim=0cm 0cm 4.5cm 0cm, clip=true]{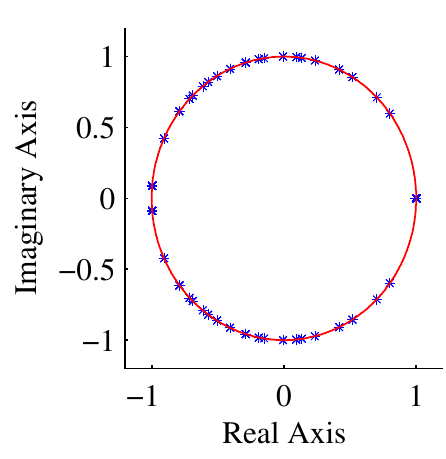}}
\caption{Example of Sec.~\ref{sec:demo}: eigenvalues of FDTD equations~\pref{reducedMNAform} for $s=1.98$ after stability has been enforced using the method of Sec.~\ref{sec:enforce}.}
\label{demo_fig2}
\end{figure}

We illustrate the proposed stability enforcement method on a simple example. We consider a 1~m $\times$ 1~m $\times$ 1~m PEC cavity discretized with a 3-D FDTD grid with cell size $\Delta = 0.\dot{1}\,{\rm m}$ along each dimension.
We let the excitation be a Gaussian pulse with a maximum frequency of 0.3~GHz, which leads to an effective $\lambda/\Delta$ of 10 at 0.3~GHz.  There are two resonant frequencies within the excitation bandwidth, one at 0.21199~GHz and one at 0.25963~GHz.  Fig.~\ref{demo_fig1} plots the eigenvalues of~\pref{MNAform} for $s=0.99$ and $s=1.98$. With $s$, we denote the CFL extension factor
\begin{equation}
	s = \frac{\Delta t}{\Delta t_{max}}\,,
	\label{eq:s}
\end{equation}
where $\Delta t_{max}$ is the maximum timestep compatible with the CFL limit~\pref{eq:CFL}. In the first case, timestep is below the CFL limit and all eigenvalues fall inside the unit circle, as shown in the left panel of Fig.~\ref{demo_fig1}. In the second case, since timestep violates the CFL constraint, some eigenvalues move into the unstable region as shown in Fig.~\ref{demo_fig1}, right panel. We therefore apply the stability enforcement procedure of Sec.~\ref{sec:enforce}, perturbing the $\K$ matrix in~\pref{MNAform}. Fig.~\ref{demo_fig2} shows the eigenvalues of the perturbed system which are all stable since they fall on the unit circle. 

In this example, due to the small size of the problem, we have enforced stability directly on the original FDTD equations. For larger problems, such as those that will be presented in Sec.~\ref{sec:results}, enforcement will be performed after the size of the problem has been reduced through model order reduction.

\section{Practical Implementation}
\label{sec:implementation}

In this Section, we discuss how the proposed method has been implemented for maximum computational efficiency.

\subsection{Complex Frequency Hopping} 

SPRIM~\cite{SPRIM}, like all order reduction methods based on moment matching, generates a reduced model accurate near a given \emph{expansion point} in the complex frequency plane. As the order of the reduced model is increased through $\tilde{N}$, the bandwidth of validity of the reduced model around the expansion point grows. In terms of eigenvalues, as $\tilde{N}$ increases, more and more eigenvalues of the original system will be matched by the reduced model eigenvalues, starting from those closer to the expansion point. By choosing the position of the expansion point, one can thus optimize the accuracy of the reduced model, and guide its convergence towards the most relevant eigenvalues of the system. 

In order to ensure accuracy over a large bandwidth, it is common practice to take multiple expansion points using the so-called ``complex frequency hopping''~\cite{CFH}. In this work, we use the following distribution of expansion points
\begin{equation}
	z_l = M e^{j 2 \pi \tfrac{l}{L} f_{max} \Delta t } \,,
	\label{eq:distrib}
\end{equation}
for $l =-L, \dots,0,\dots, L$. This formula places $2 L + 1$ expansion points along a circular arc of radius $M$ centered at the origin of the complex plane. The magnitude of the points $M$ is chosen to be slightly higher than 1, and $M=1.1$ will be used in all numerical examples of Sec.~\ref{sec:results}. With this distribution, one expansion point is  always placed near $z=1$. Through this expansion point we capture the low-frequency response of the system, since $z=1$ corresponds to static conditions. The other expansion points are placed near the unit circle up to the maximum frequency of interest $f_{max}$. With them, we capture the eigenvalues along the unit circle, starting from those that are in magnitude close to one. These eigenvalues have indeed a significant impact on the system response, since they correspond to weakly-damped modes. Eigenvalues well inside the unit circle are instead highly-damped, and their contribution to the system response quickly fades away. It has been experimentally determined that distribution~\pref{eq:distrib} significantly improves accuracy for a fixed reduced model size. An additional benefit is the reduction of the Gram-Schmidt orthogonalizations required to generate the Krylov subspace~\cite{SPRIM,PRIMA}.

\subsection{Linear System Solution}

A linear system must be solved for each new moment generated with the Arnoldi process used to generate~\pref{eq:V}. The  system is in the form
\begin{equation}
\left[({\bf R} - {\bf F}) + z_l ({\bf R} + {\bf F}) \right]
\begin{bmatrix}
\vect{x}_1\\
\vect{x}_2
\end{bmatrix} 
=
\begin{bmatrix}
\vect{b}_1\\
\vect{b}_2
\end{bmatrix} 
\label{linear_system_solution}
\end{equation}
where $z_l$ is the current expansion point.
When $z_l = 0$, since $({\bf R} - {\bf F})$ is upper triangular, the system can be solved very efficiently. For $z_l \neq 0$, the system can be solved with one LU decomposition~\cite{Gol96} in 2-D and small 3-D cases, similarly to what done in~\cite{Subcell} and~\cite{SecondOrder}. For large 3-D cases, iterative methods must be used, and we adopted the conjugate gradient squared method~\cite{Bar94} available in Matlab. We determined, through the test cases of Sec.~\ref{sec:results}, that with a normalized residue limit of 10$^{-4}$, the iterative solver leads to results comparable to those obtained with a direct solver.

In solving~\pref{linear_system_solution}, we also exploit the 2x2 block structure of $\matr{R}$ and $\matr{F}$. Let us denote the four blocks of the system matrix as
\begin{equation}
\begin{aligned}
({\bf R} - {\bf F}) + z_l ({\bf R} + {\bf F}) &=
\begin{bmatrix} 
  {\bf A}_{11} & {\bf A}_{12} \\
  {\bf A}_{21} & {\bf A}_{22} \\
\end{bmatrix} \\
\end{aligned}
\end{equation}
where ${\bf A}_{11}$ and ${\bf A}_{22}$ are diagonal matrices.  Using the Schur complement~\cite{Boy94}, we first solve for $\vect{x}_1$ in~\pref{linear_system_solution}
\begin{align}
({\bf A}_{11}-{\bf A}_{12}{\bf A}_{22}^{-1}{\bf A}_{21}) \vect{x}_1  &= \vect{b}_1 - {\bf A}_{12}{\bf A}_{22}^{-1} \vect{b}_2
\label{reduced_linear_system_solution1}
\end{align}
and then solve for $\vect{x}_2$
\begin{align}
\vect{x}_2 &= {\bf A}_{22}^{-1} (\vect{b}_2 - {\bf A}_{21} \vect{x}_1)
\label{reduced_linear_system_solution2}
\end{align}
which can be done very quickly since $\matr{A}_{22}$ is diagonal. With the implementation discussed in this section, we were able to apply the proposed method to 3-D simulations of practical relevance, with more than one million unknowns.

\section{Numerical Results}
\label{sec:results}

\subsection{2-D and 3-D Cavities}
\label{sec:cavity}
The proposed method to reduce FDTD equations and extend the CFL limit was implemented in MATLAB, and applied to several test structures. First, we consider two empty cavities with PEC walls, one in a 2-D setting, and the other one in a 3-D setting. The sidelength of the cavity is 1~m in both cases.  A single source and probe are placed to capture all resonant frequencies within the interested range.  The input is a Gaussian pulse with a bandwidth of 0.5~GHz.  The proposed method has been compared against 4 other approaches in the literature: an implicit integration of Maxwell's equations combined with model order reduction~\cite{Subcell}, ADI-FDTD~\cite{ADI-FDTD}, spatial filtering~\cite{SpatialFiltering}, and the reduction algorithm of~\cite{DJiao,DJiao2014}.  We investigated their overall run time, accuracy, and approximate numerical dispersion below and above the CFL limit.  Analytical resonant frequencies were calculated and used as accuracy metric. All simulations were run for 10,000 timesteps in order to achieve sufficient resolution in the frequency domain. They were also run for $10^6$ timesteps to determine their late time stability. The most relevant simulation settings are summarized in Table~\ref{cavity_table1}.

\begin{table}
\centering
	\caption{Cavities of Sec.~\ref{sec:cavity}: simulation parameters.}
    \begin{tabular}{ | l | l | l |}
    \hline
    Property & 2-D Cavity & 3-D Cavity \\ \hline \hline
    Time Steps & 10,000 & 10,000 \\ \hline
    
    x-axis Cells & 101 & 51 \\ \hline
    y-axis Cells & 101 & 51 \\ \hline
    z-axis Cells & - & 51 \\ \hline
    $\Delta x$, $\Delta y$, $\Delta z$ & 1 cm & 2 cm \\ \hline
    
    $\lambda / \Delta$ at 0.5~GHz & 60 & 30 \\ \hline
    System Size $N$ & 30,200 & 795,906 \\ \hline
    
    \end{tabular}
    \label{cavity_table1}
\end{table} 

For the proposed method, we set the order of the reduced model to 80. Five expansion points were used, distributed in the complex plane according to~\pref{eq:distrib} with $M=1.1$ and $f_{max} = 0.5\,{\rm GHz}$. A direct linear system solver (LU decomposition) was used in the 2-D case, while the iterative conjugate gradient squared method was used for the 3-D case. 
The settings of the other methods were selected in order to obtain a comparable accuracy. A reduced order of 40 was used for~\cite{Subcell}, while the method of Gaffar and Jiao required 500~timesteps to accurately identify the important system eigenvalues.  The weighting coefficient, $\epsilon_1$, of Gaffar and Jiao was set at 10$^{-3}$.

\begin{figure}[t]
\centering
{\includegraphics[scale=.8]{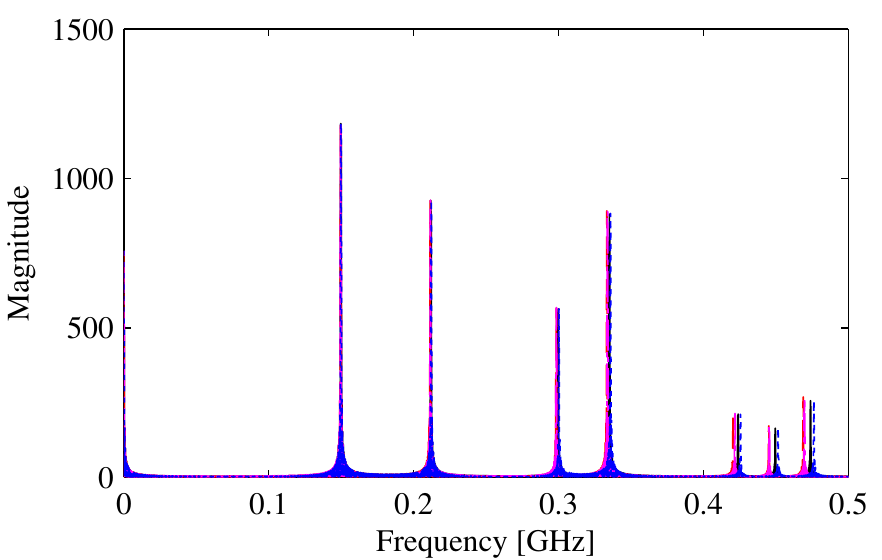}}
\caption{2-D Cavity of Sec.~\ref{sec:cavity}: frequency response of the cavity obtained from the results of Yee's FDTD ({\color{black} \st{$\;\;\;\;$}}), the implicit method of~\cite{Subcell}  ({\color{red} $--$}) and ADI-FDTD  ({\color{magenta} -$-$-}). The proposed method, spatial filtering~\cite{SpatialFiltering}, and the method of Gaffar and Jiao~\cite{DJiao} gave the same results and are depicted with a single curve ({\color{blue} -\,\,-\,\,-}). Yee's FDTD was run below the CFL limit ($s=0.99$). All other methods were run above the CFL limit ($s=4.95$).}
\label{cavity_fig1}
\vspace{0.2cm}
\includegraphics[scale=0.47]{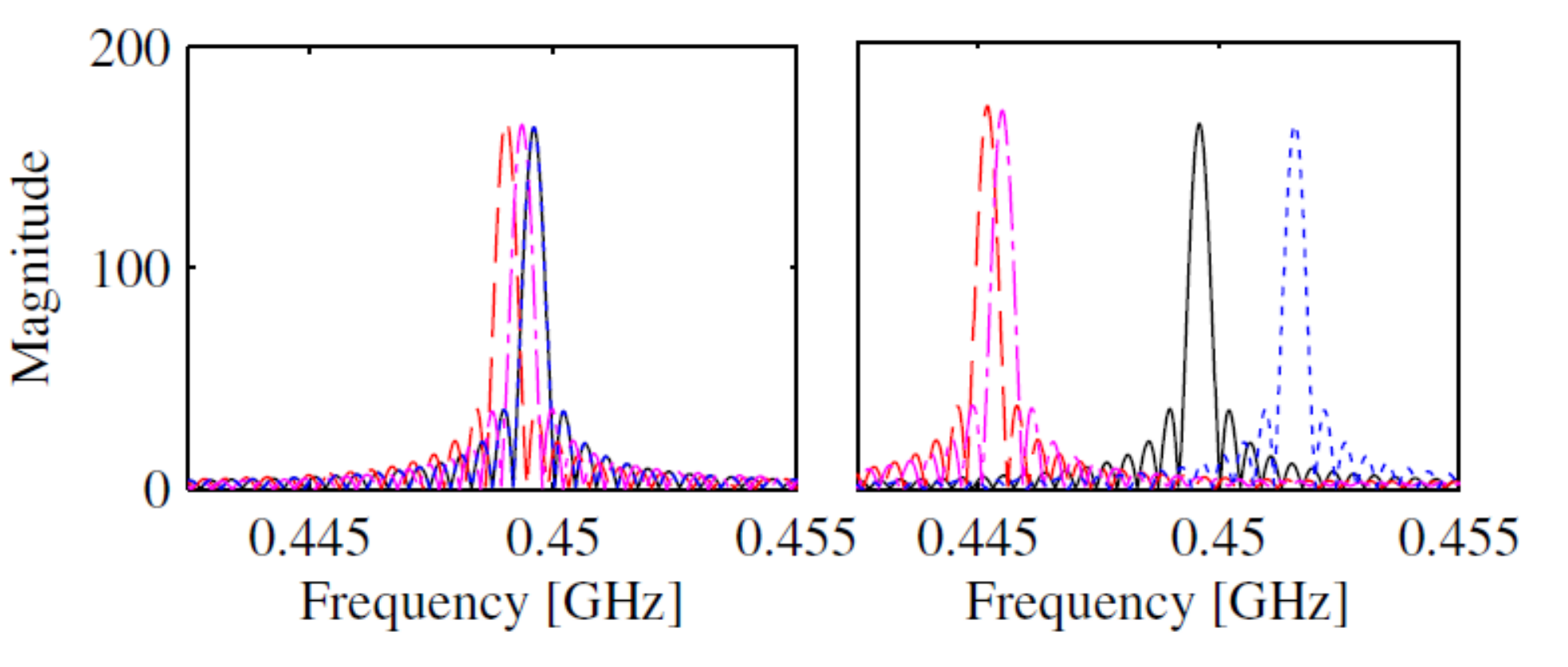}
\caption{As in Fig.~\ref{cavity_fig1}, but with focus on the TM$_{30}$ resonance. The different methods were run at $s=0.99$ (left panel) and at $s=4.95$ (right panel). Yee's FDTD was run in both cases at $s=0.99$.}
\label{cavity_fig2}
\end{figure}

\begin{figure}[t]
\centering
{\includegraphics[scale=.9]{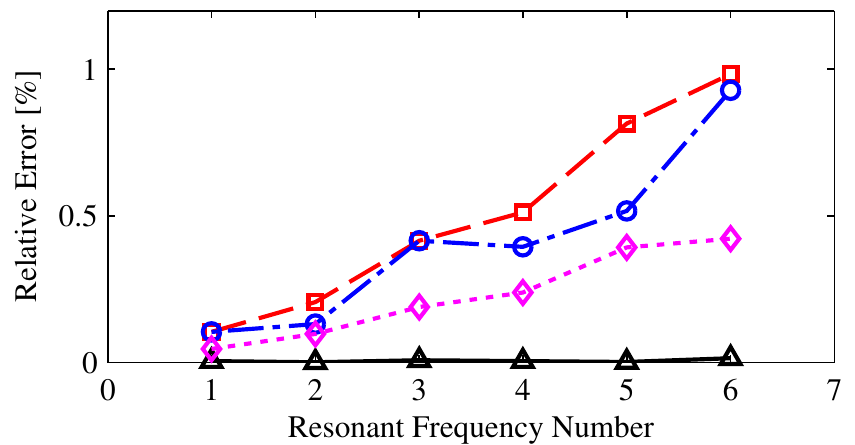}}
\caption{2-D Cavity of Sec.~\ref{sec:cavity}: relative error on the first 6 resonant frequencies obtained with Yee's FDTD run at $s=0.99$(\st{$\;\bigtriangleup\;$}), the implicit method of~\cite{Subcell} ({\color{red} \st{$\;\Box\;$}}), and ADI-FDTD ({\color{blue} \st{$\;\ominus\;$}}). The proposed method, spatial filtering~\cite{SpatialFiltering}, and the method of Gaffar and Jiao~\cite{DJiao} gave the same results and are depicted with a single curve ({\color{magenta} \st{$\;\Diamond\;$}}). Yee's FDTD was run below the CFL limit ($s=0.99$). All other methods were run above the CFL limit ($s=4.95$). Gold standard: analytical formulas.}
\label{cavity_fig3}
\end{figure}

\begin{figure}[t]
\centering
{\includegraphics[scale=.9]{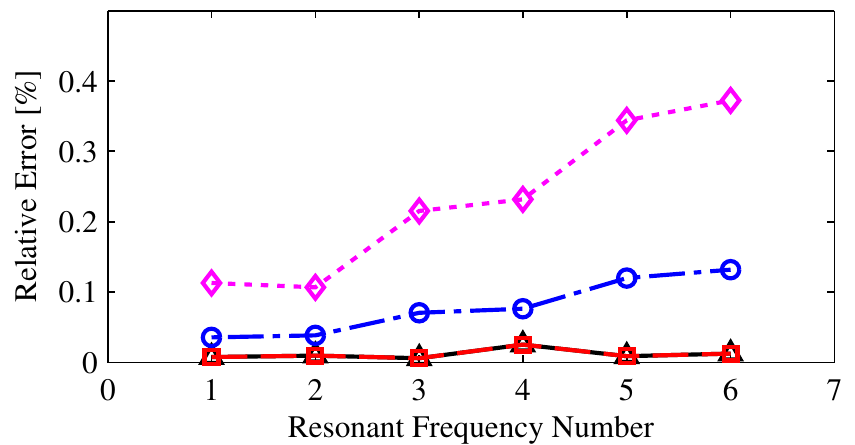}}
\caption{3-D Cavity of Sec.~\ref{sec:cavity}: relative error on the first 6 resonant frequencies. Yee's FDTD (\st{$\;\bigtriangleup\;$}) at CFL = 0.99 and proposed method at $s=0.99$ ({\color{red} \st{$\;\Box\;$}}), $s=1.98$ ({\color{blue} \st{$\;\ominus\;$}}), and $s=2.97$ ({\color{magenta} \st{$\;\Diamond\;$}}). Gold standard: analytical formulas.}
\label{cavity_fig4}
\end{figure}

We investigate the accuracy of the different methods by looking at resonant frequencies. The frequency response of the 2-D cavity is depicted in Fig.~\ref{cavity_fig1}, where we can observe that all methods accurately capture the resonant frequencies of the structure, even when run above the CFL limit. Only small deviations can be observed in the highest resonances. A zoom on the resonance at $0.45\,{\rm GHz}$ is provided in Fig.~\ref{cavity_fig2}. The small increase in dispersion due to the CFL extension can be observed by comparing the two panels of Fig.~\ref{cavity_fig2}. Figure~\ref{cavity_fig3} shows the relative error on the first resonances for the different methods, which may be attributed to numerical dispersion.  When run above the CFL limit, all methods introduce some additional dispersion with respect to Yee's FDTD run below the CFL limit. It can be observed that explicit methods (proposed method, \cite{SpatialFiltering} and \cite{DJiao}) introduce less dispersion than implicit alternatives~\cite{ADI-FDTD,Subcell}.
 Figure~\ref{cavity_fig4} refers to the 3-D cavity and shows the relative error on the first 6 resonant frequencies obtained with Yee's FDTD and the proposed method run at different CFL extension factors, which remains well below 1\% in all cases.

Table~\ref{cavity_table2} shows the simulation time break down for the 2-D case. 
Below the CFL limit ($s=0.99$), the proposed method ensures a speed-up of 4.32X with respect to FDTD, thanks to the reduced size of the generated model, which makes its solution very cheap. The proposed method is also faster than the other tested methods. CFL extension techniques such as ADI-FDTD and spatial filtering are not necessary for $s=0.99$, but have been included to illustrate runtime scaling.  When run above the CFL limit, at $s=4.95$, all methods delivered stable results and achieved a higher speed-up with respect to FDTD. The proposed method offers a speed-up of about 5X with respect to FDTD, which is comparable to the speed-up obtained with the method of Gaffar and Jiao~\cite{DJiao}. 

\begin{table}
\centering
    \caption{2-D Cavity of Sec.~\ref{sec:cavity}: execution time breakdown for the different methods. All times are in seconds.}
    \begin{tabular}{ | p{1.3cm} | p{0.4cm} | l | l | l || l || p{0.6cm} |}
    \hline
    Case & Size & Setup & MOR & Run & Total & Speed- up \\ \hline
    2-D FDTD 	& & 0.01 &  & 4.70 & 4.71 & - \\ \hline
    $s = 0.99$ 	& & & & & & \\ \hline
	Implicit MOR~\cite{Subcell} 				& 40 	& 0.19 & 1.27 & 0.13 	& 1.60 	& 2.94\\ \hline
	ADI-FDTD~\cite{ADI-FDTD} 					& - 		& 0.33	& - 	& 27.87 	& 28.20 	& 0.16\\ \hline
	Spatial Filtering~\cite{SpatialFiltering} 	& - 		& 0.01	& - 	& 16.24 	& 16.25 	& 0.29\\ \hline
	Gaffar and Jiao~\cite{DJiao} 				& 144 	& 0.10 & 0.84 & 0.47 	& 1.41  & 3.34\\ \hline
	Proposed 									& 80 	& 0.08 & 0.74 & 0.25 	& 1.09 	& 4.32\\ \hline \hline 
	$s = 4.95$ 	& & & & & & \\ \hline
	Implicit MOR~\cite{Subcell} 				& 40 	& 0.18 & 1.21 & 0.02 	& 1.43 	& 3.29\\ \hline
	ADI-FDTD~\cite{ADI-FDTD} 					& - 		& 0.34	& - 	& 5.39 	& 6.10 	& 0.77\\ \hline
	Spatial Filtering~\cite{SpatialFiltering} 	& - 		& 0.00	& - 	& 3.38 	& 3.39 	& 1.38\\ \hline
	Gaffar and Jiao~\cite{DJiao} 				& 52 	& 0.10 & 0.84 & 0.04 	& 0.98 	& 4.80\\ \hline
	Proposed  									& 80 	& 0.08 & 0.73 & 0.06 	& 0.88 	& 5.37\\ \hline
    \end{tabular}
    \label{cavity_table2}
\end{table} 

Table~\ref{cavity_table3} shows the simulation time break down for the proposed method and FDTD for the 3-D case. It can be observed that the proposed method demonstrates a significant speed-up over standard FDTD.

\begin{table}
\centering
    \caption{3-D Cavity of Sec.~\ref{sec:cavity}: execution time breakdown for FDTD and the proposed method for different CFL extension factors. All times are in seconds.}
    \begin{tabular}{ | p{1.3cm} | p{0.4cm} | l | l | l || l || p{0.6cm} |}
    \hline
    Case & Size & Setup & MOR & Run & Total & Speed- up \\ \hline
	3-D FDTD & & 2.24 &  &  344.28 & 346.52 & - \\ \hline
	Proposed (s = 0.99) & 80 & 3.56 & 99.23 & 0.25 & 103.04 & 3.36\\ \hline
	Proposed (s = 1.98) & 80 & 3.56 & 99.23 & 0.12 & 102.91 & 3.36\\ \hline
	Proposed (s = 2.97) & 80 & 3.56 & 99.23 & 0.09 & 102.88 & 3.36\\ \hline
    \end{tabular}
    \label{cavity_table3}
\end{table}

\subsection{2-D Waveguide with Irises}
\label{sec:waveguide}

\begin{figure}[t]
\centering
{\includegraphics[scale=.9]{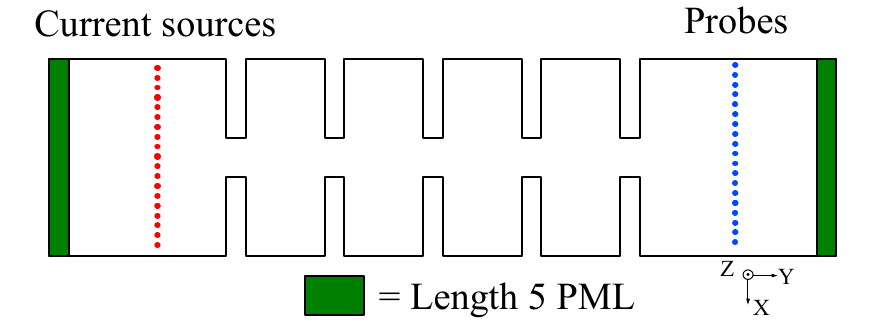}}
\caption{Layout of the waveguide considered in Sec.~\ref{sec:waveguide}. For readability, axes are not in scale.}
\label{waveguide_fig1}
\end{figure}

A 2-D waveguide filter operating in the TM mode of size 5~cm $\times$ 50~cm is discretized into a 41 $\times$ 401 mesh with $\Delta x$ = $\Delta y$ = 1.25~mm. The layout of the waveguide is shown in Fig.~\ref{waveguide_fig1}. A  Gaussian current line source with bandwidth of 3~GHz is placed at one end of the waveguide, while a line probe is placed on the other end.  
The waveguide is filled with a dielectric material with $\epsilon_r=2.5$.  Five irises (length: 1.25~cm, aperture size: 1~cm, separation: 5~cm) are evenly placed within the waveguide.  The waveguide is terminated at both ends on a 4th-order matched absorber with thickness of 5~cells.
A matched absorber is used for simplicity, although we have shown in~\cite{IMS} that a split PML with auxiliary equations can also be used.   The minimum $\lambda/\Delta$ is 80 at 3~GHz.   The original system size is 48,440, and the size of the reduced model generated with the proposed algorithm is 200.  Due to the wide band of the excitation, 5 expansion points were placed on the complex plane according to~\pref{eq:distrib} with $M= 1.1$ and $f_{max} = 3\,{\rm GHz}$. A direct solver (LU decomposition) was used to solve~\pref{linear_system_solution}.  FDTD and the proposed technique were run for 20,000 timesteps, in order to allow for the the input power to dissipate in the structure.

\begin{figure}[t]
\centering
{\includegraphics[scale=.95]{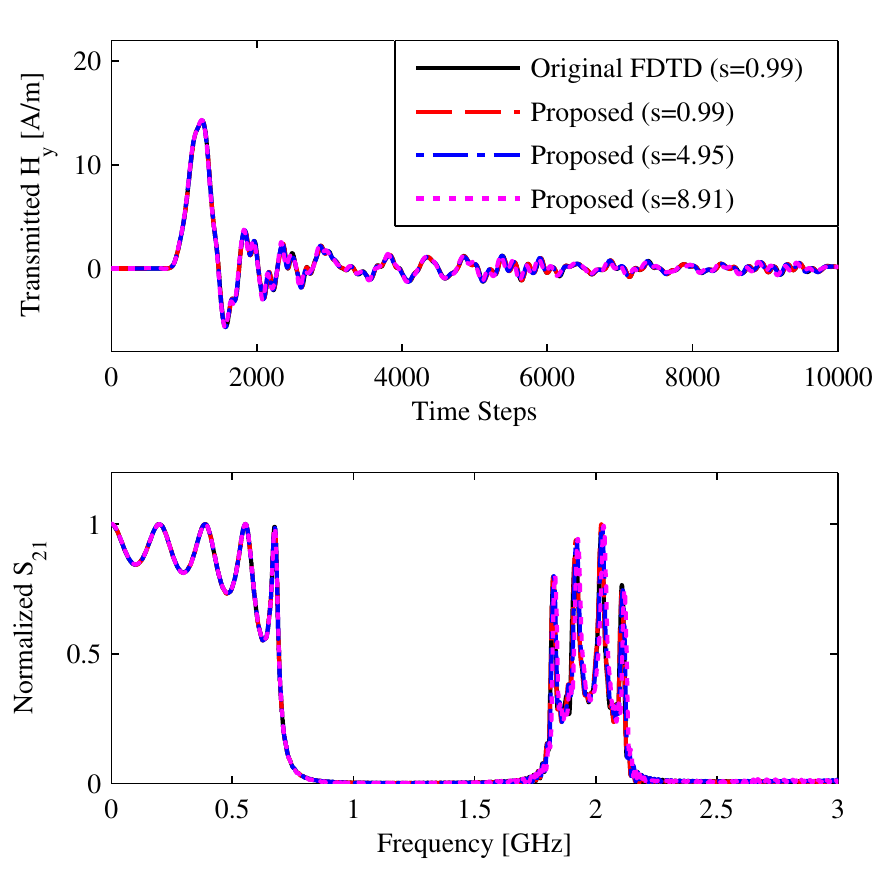}}
\caption{Waveguide of Sec.~\ref{sec:waveguide}: magnetic field at the probes (top) and $S_{21}$ transmission coefficient (bottom) for Yee's FDTD at CFL = 0.99 and for the proposed method at $s=0.99$, 4.95, and 8.91.}
\label{waveguide_fig2}
\end{figure}

Fig.~\ref{waveguide_fig2} shows the simulation results for both standard FDTD and the proposed method run below and above the CFL limit. Top panel depicts the 
magnetic field received by the probes, and shows the stability of the proposed method, even when run 5 times above the CFL limit. Bottom panel gives the $S_{21}$ transmission coefficient over frequency. Excellent accuracy in both time and frequency domain may be observed. Timing results for FDTD and the proposed method are summarized in Table~\ref{2D_waveguide_table2}. The proposed method substantially accelerates the analysis of the waveguide, giving a speed up of 11X at the CFL limit ($s=0.99$), and of almost 16X when run with a timestep which is 9~times larger than the CFL limit ($s=8.91$). These results confirm the advantages of the proposed method, which combines model order reduction and eigenvalue perturbation to accelerate FDTD simulations beyond the CFL limit.

\begin{table}
\centering
    \caption{Waveguide of Sec.~\ref{sec:waveguide}: execution time breakdown for Yee's FDTD and the proposed method at refined CFL numbers. All times are in seconds.}
    \begin{tabular}{ | l | l | l | l | l || l || l |}
    \hline
    Case & Size  & Setup & MOR & Run & Total & Speedup \\ \hline
    FDTD & & 0.01 &  & 37.17 & 37.18 & - \\ \hline
	s = 0.99 & 200 & 0.7 & 1.36 & 1.25 & 3.31 & 11.2\\ \hline
	s = 4.95 & 200 & 0.7 & 1.36 & 0.25 & 2.31 & 16.0\\ \hline
	s = 8.91 & 200 & 0.7 & 1.36 & 0.14 & 2.20 & 16.9\\ \hline
    \end{tabular}
    \label{2D_waveguide_table2}
\end{table}

\subsection{3-D Focusing Metascreen}
\label{sec:propagation}

The proposed method is applied to the focusing metascreen structure first proposed in~\cite{SlitFocusing-1} and subsequently investigated using FDTD in~\cite{SlitFocusing-2}.  This test case involves the transmission of a plane wave through a metallic screen with a central slot for focusing onto an image plane.  
The 3-D simulation domain is of size 61 $\times$ 61 $\times$ 61 with $\Delta x = \Delta y = \Delta z = 0.3\,{\rm mm}$, and is terminated on all sides with 4th-order, 5-cell matched absorbers.  A plane of uniform sinusoidal sources at 10~GHz is placed on one side of the PEC screen, which is one-cell thick. The screen has a single focusing slot of size 13.2~mm $\times$ 1.2~mm. Probes are placed on the other side of the screen along the centre axis of the image plane, at a distance of $0.15\lambda$ from the screen.  A very high $\lambda / \Delta$ ratio of 100 is required due to the resonating nature of the slot. The fine mesh makes the size of the original FDTD equations quite high ($N = 1,361,886$). 
Due to the single-frequency excitation, only a pair of expansion points were used, given by 
\begin{equation}
	z_{1,2} = M e^{\pm j 2 \pi f_{max} \Delta t } \,,
\end{equation}
with  $M=1.1$ and $f_{max} = 10\,{\rm GHz}$. The proposed method was used to generate a reduced model of order 40 and simulations were run for 10,000 timesteps until a steady state was reached on the image plane.

\begin{figure}[t]
\centering
{\includegraphics[scale=.85]{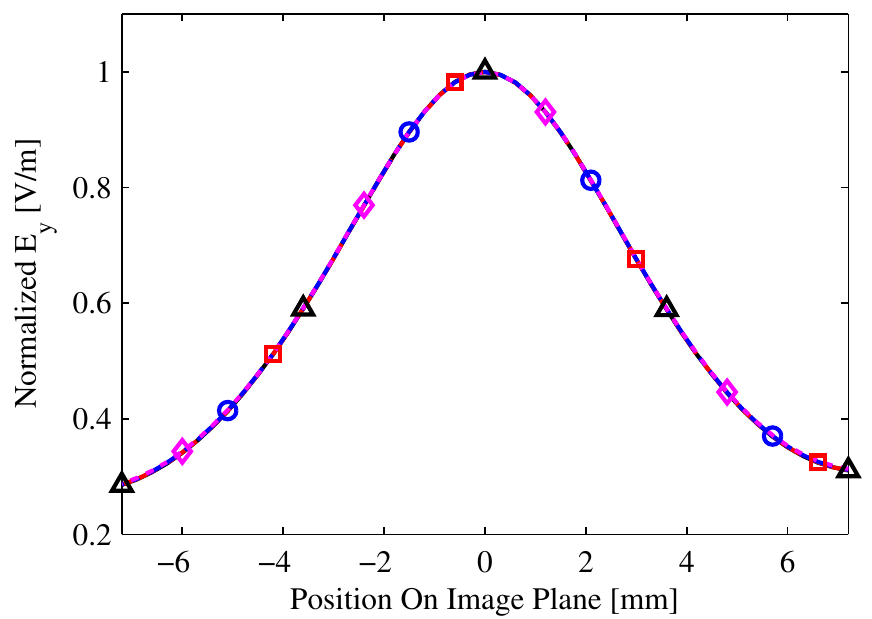}}
\caption{Focusing metascreen of Sec.~\ref{sec:propagation}: average amplitude of the electric field E$_y$ on the image plane.  Comparison of Yee's FDTD (\st{$\;\bigtriangleup\;$}) at $s=0.99$ against the proposed method at $s=0.99$ ({\color{red} \st{$\;\Box\;$}}), $s=2.97$ ({\color{blue} \st{$\;\ominus\;$}}), and $s=4.95$ ({\color{magenta} \st{$\;\Diamond\;$}}).}
\label{focusing_fig2}
\end{figure}

Fig.~\ref{focusing_fig2} compares the electric field on the image plane calculated with the proposed method and with standard FDTD. An excellent match can be observed, even when the proposed method is run at 5 times the maximum timestep allowed in conventional FDTD. Table~\ref{3D_focusing_table2} shows the simulation time breakdown for standard FDTD and the proposed method, which leads to a speed up of 4.28X. In this case, the extension of the CFL limit has a small influence on the total solution time for the proposed method, since the reduction step dominates the solution of the reduced model due to the fairly large size of the problem. Future investigations will focus on improving the efficiency of the reduction process for large-scale problems.

\begin{table}
\centering
    \caption{Focusing metascreen of Sec.~\ref{sec:propagation}: execution time breakdown for Yee's FDTD and proposed method at refined CFL numbers. All times are in seconds.}
    \begin{tabular}{ | l | l | l | l | l || l || l |}
        \hline
    Case & Size & Setup & MOR & Run & Total & Speedup \\ \hline
    	3-D FDTD & & 3.8 &  & 561.4 & 565.2 & - \\ \hline
    	s = 0.99 & 40 & 5.6 & 125.9 & 0.23 & 131.8 & 4.29\\ \hline
    	s = 2.97 & 40 & 5.6 & 125.9 & 0.07 & 131.6 & 4.29\\ \hline
    	s = 4.95 & 40 & 5.6 & 125.9 & 0.05 & 131.6 & 4.29\\ \hline
    \end{tabular}
    \label{3D_focusing_table2}
\end{table}

\subsection{3-D Microstrip Filter}
\label{sec:filter}

\begin{figure}[t]
\centering
{\includegraphics[scale=0.7]{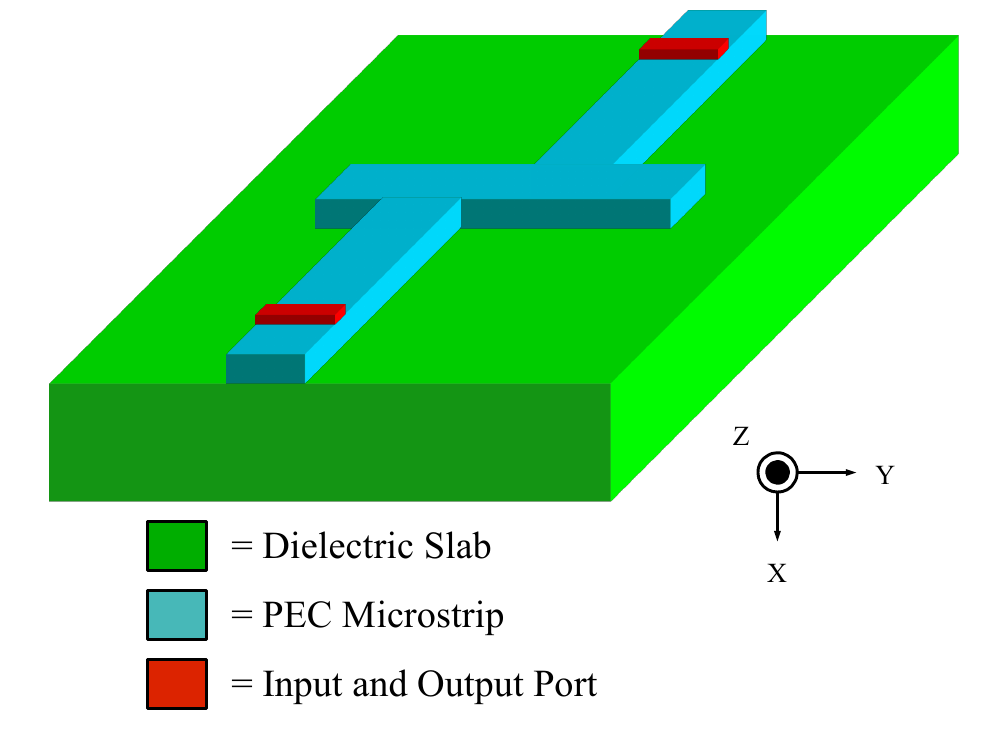}}
\caption{Microstrip filter example of Sec.~\ref{sec:filter}: layout.}
\label{microstrip_fig1}
\end{figure}

The final test case is an application of the proposed method to a 3-D multi-port microstrip filter from~\cite{microstrip}. The structure is shown in Fig.~\ref{microstrip_fig1}. The simulation domain is of size 81 $\times$ 91 $\times$ 14 cells, with $\Delta x = \Delta y = 0.4\,{\rm mm}$ and $\Delta z = 0.2\,{\rm mm}$. A one-cell thick PEC microstrip rests on a dielectric substrate with $\epsilon_r=2$ and thickness of 3 cells.  The PEC microstrip is 7-cells wide (2.8~mm) and two ports are placed at its ends.  
The domain is terminated on 5 sides with 5-cell 4th-order matched absorbers.  A PEC wall is used for the 6-th side to model the ground plane.  The size of the original FDTD equations~\pref{MNAform} is 619,164. The simulation utilizes uniform line sources and probes.  A Gaussian pulse of 20~GHz bandwidth is used to extract the $S_{21}$ and $S_{11}$ parameters.
A reduced model~\pref{reducedMNAform} of order 100 was generated using 5 expansion points~\pref{eq:distrib} with $M= 1.1$ and $f_{max} = 20\,{\rm GHz}$. Simulations were run for 8,000 timesteps, when most of the input power was dissipated.
 
\begin{figure}[t]
\centering
{\includegraphics[scale=1]{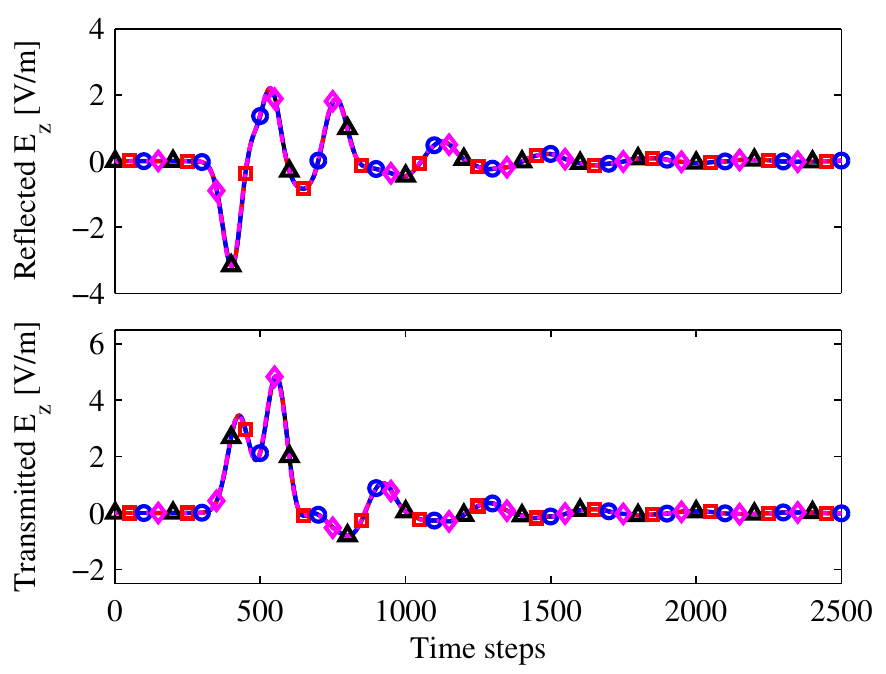}}
\caption{Microstrip filter of Sec.~\ref{sec:filter}: time domain reflected (top) and transmitted (bottom) waveforms computed with Yee's FDTD (\st{$\;\bigtriangleup\;$}) at $s= 0.99$ and with the proposed method at $s=0.99$ ({\color{red} \st{$\;\Box\;$}}), $s=2.97$ ({\color{blue} \st{$\;\ominus\;$}}), and $s=4.95$ ({\color{magenta} \st{$\;\Diamond\;$}}).}
\label{microstrip_fig2}
\end{figure}

\begin{figure}[t]
\centering
{\includegraphics[scale=.9]{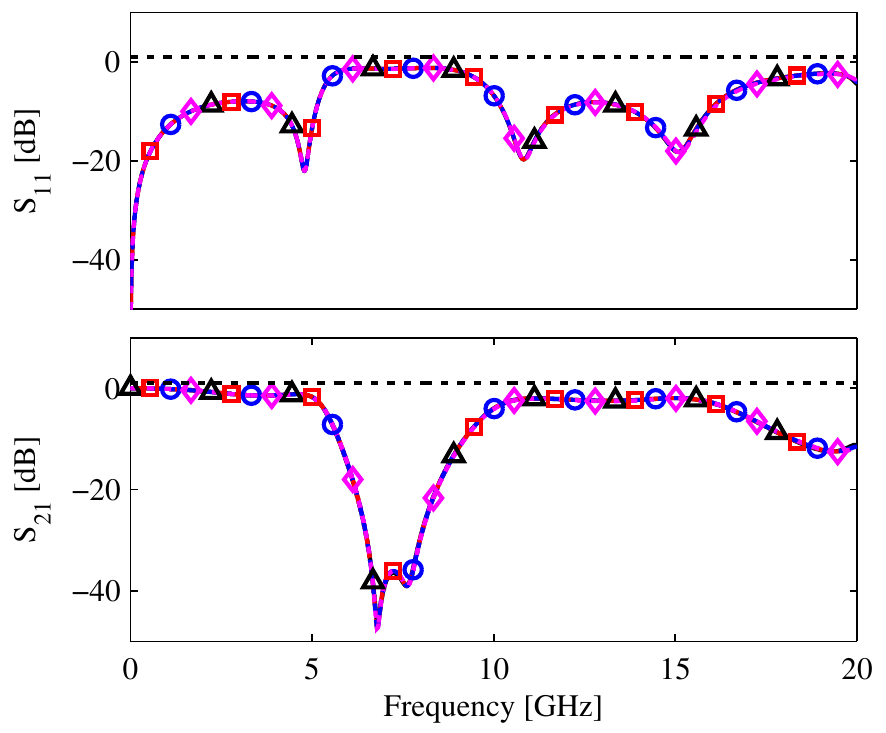}}
\caption{Microstrip filter of Sec.~\ref{sec:filter}: S$_{11}$ (top) and S$_{21}$ (bottom) parameters obtained with Yee's FDTD (\st{$\;\bigtriangleup\;$}) at $s= 0.99$ and with the proposed method at $s=0.99$ ({\color{red} \st{$\;\Box\;$}}), $s=2.97$ ({\color{blue} \st{$\;\ominus\;$}}), and $s=4.95$ ({\color{magenta} \st{$\;\Diamond\;$}}).}
\label{microstrip_fig3}
\end{figure} 
 
 Fig.~\ref{microstrip_fig2} depicts the time domain reflected and transmitted waveforms, while Fig.~\ref{microstrip_fig3} plots the $S_{11}$ and $S_{21}$ parameters extracted from the time domain analysis. The reflected and transmitted waveforms generated with the proposed method, even at 5 times the maximum timestep allowed by the CFL limit,  are indistinguishable from those obtained with Yee's FDTD at $s=0.99$. These results further confirm the excellent accuracy of the proposed technique and its stable behavior even at timesteps significantly higher than the CFL limit.
 
Finally, table~\ref{3D_microstrip_table2} shows the simulation time breakdown for the proposed method compared to Yee's FDTD at various CFL extension factors. Also in this case we observe a speed-up with respect to standard FDTD, with essentially no loss of accuracy.

\begin{table}
\centering
    \caption{Microstrip filter of Sec.~\ref{sec:filter}: execution time breakdown for Yee's FDTD and the proposed method. All times are in seconds.}
    \begin{tabular}{ | l | l | l | l | l || l || l |}
    \hline
    Case & Size & Setup & MOR & Run & Total & Speedup \\ \hline
	3-D FDTD & & 1.7 &  & 233.2 & 234.9 & - \\ \hline
	s = 0.99 & 100 & 2.9 & 84.7 & 0.46 & 88.1 & 2.71\\ \hline
	s = 2.97 & 100 & 2.9 & 84.7 & 0.15 & 87.8 & 2.72\\ \hline
	s = 4.95 & 100 & 2.9 & 84.7 & 0.09 & 87.7 & 2.72\\ \hline
    \end{tabular}
    \label{3D_microstrip_table2}
\end{table}

\section{Conclusion}
We proposed a new way to perform stable FDTD simulations beyond the CFL limit. A new approach to the order reduction of FDTD equations is first developed. Differently from previous works, our approach works directly in the discrete time domain. It preserves the structure of FDTD equations and, for timesteps below the CFL limit, guarantees late-time stability by construction. Then, we show how it is also possible to enforce the stability of the reduced model above the CFL limit. The proposed method can handle non-uniform grids, losses, and non-homogeneous materials. Numerical tests on structures based on microstrips, waveguides and resonant cavities were presented to demonstrate the superior efficiency and the excellent accuracy of the technique. 

\bibliographystyle{IEEEtran}
\bibliography{IEEEabrv,bibliography}

\end{document}